\documentclass[conference]{IEEEtran}
\IEEEoverridecommandlockouts
\usepackage{cite}
\usepackage{amsmath,amssymb,amsfonts}
\usepackage{algorithmic}
\usepackage{graphicx}
\usepackage{textcomp}
\usepackage{xcolor}
\def\BibTeX{{\rm B\kern-.05em{\sc i\kern-.025em b}\kern-.08em
    T\kern-.1667em\lower.7ex\hbox{E}\kern-.125emX}}
    
\usepackage{graphicx}
\usepackage{tikz}
\usepackage{caption}
\usepackage{subcaption}
\usepackage{tikz}
\usetikzlibrary{spy,calc}
\usepackage{verbatim}
\usepackage{booktabs,array} 
\usepackage{url}

\usepackage{siunitx}
\usepackage{microtype}
\usepackage{cite}
\usepackage{amsmath,amsfonts,amssymb,amsthm}
\usepackage{mathtools}

\usepackage[ruled,vlined]{algorithm2e}

\usepackage{pgfplots}
\pgfplotsset{compat=newest, legend style={at={(1,0.05)},anchor=south east}
}

\usepackage{hyperref}

\hypersetup{hidelinks,
backref=true,
pagebackref=true,
hyperindex=true,
breaklinks=true,
colorlinks=true,
urlcolor=blue,
bookmarks=true}

\usepackage{cleveref}

\crefname{section}{Sec.}{Sections}
\crefname{figure}{Fig.}{Figure}
\crefname{table}{Tab.}{Table}
\crefname{equation}{Eq.}{Equation}

\newcolumntype{L}[1]{>{\raggedright\arraybackslash}p{#1}}

\makeatletter
\DeclareRobustCommand\onedot{\futurelet\@let@token\@onedot}
\def\@onedot{\ifx\@let@token.\else.\null\fi\xspace}
\makeatother

\newcommand{\cf}{cf\onedot}
\newcommand{\ie}{i.\,e.,\xspace}

\makeatletter
\newcommand{\thickhline}{%
    \noalign {\ifnum 0=`}\fi \hrule height 1pt
    \futurelet \reserved@a \@xhline
}
\usepackage{datetime}

\makeatletter
\renewcommand\footnoterule{%
  \kern-3\p@
  \hrule\@width.4\columnwidth
  \kern2.6\p@}
  \makeatother

\begin{document}

\title{An Empirical Study on Text-Independent Speaker Verification based on the GE2E Method}

\author{\IEEEauthorblockN{Soroosh Tayebi Arasteh\IEEEauthorrefmark{1}\IEEEauthorrefmark{2}}

\\
\IEEEauthorblockA{\IEEEauthorrefmark{2}Pattern Recognition Lab, Friedrich-Alexander-Universität Erlangen-N\"urnberg, Erlangen, Germany}
\IEEEauthorblockA{\IEEEauthorrefmark{2}Department of Pediatrics, Harvard Medical School, Boston, MA, USA}
{\texttt{soroosh.arasteh@fau.de}}}

\maketitle
\thispagestyle{plain}
\pagestyle{plain}

\begin{abstract}
While many researchers in the speaker recognition area have started to replace the former classical state-of-the-art methods with deep learning techniques, some of the traditional i-vector-based methods are still state-of-the-art in the context of text-independent speaker verification. Google's Generalized End-to-End Loss for Speaker Verification (GE2E), a deep learning-based technique using long short-term memory units, has recently gained a lot of attention due to its speed in convergence and generalization. In this study, we aim at further studying the GE2E method and comparing different scenarios in order to investigate all of its aspects. Various experiments including the effects of a random sampling of test and enrollment utterances, test utterance duration, and the number of enrollment utterances are discussed in this article. Furthermore, we compare the GE2E method with the baseline state-of-the-art i-vector-based methods for text-independent speaker verification and show that it outperforms them by resulting in lower error rates while being end-to-end and requiring less training time for convergence.
\end{abstract}



\section{Introduction}


Speaker recognition (SR) is the task of recognizing the speaker's identity based on their voice. It is a very active research area with notable applications in various fields such as biometric authentication, forensics, security, and speech recognition, which has contributed to steady interest towards this discipline~\cite{bookbeigi}. Moreover, SR has become popular technology for remote authentication, especially in the advancement of telecommunications and networking~\cite{628714}.
Human speech is one of the most complex natural signals and contains a lot of information, which makes it unique for every person and enables us to create SR systems based on those properties. 

Speaker verification (SV) and speaker identification (SI) are two important subtasks of SR.
SV is the task of authenticating a person's claimed identity as genuine or imposter. SI on the other hand, is the task of identifying an unknown person's identity from a pool of known speakers. Together with SV and SI, SR is the process of identifying an unknown speaker's identity in the general case, by first verifying and then identifying.

The SV process can generally be divided into three steps of training, enrollment, and evaluation~\cite{365379}. In training stage, the speaker-specific features are extracted to create a background model for the speaker representation using the available signals. In the enrollment phase, using the background model, which is the trained network in the case of deep learning (DL) techniques, speech utterances are utilized to create the speaker models. Finally, in the evaluation step, test speaker models are created by feeding the test utterances to the background model. They are compared to already registered speaker models in order to check the similarity between them.
Furthermore, depending on the restrictions of the utterances used for enrollment and verification, SV models usually fall into one of the two categories: text-dependent speaker verification (TDSV) and text-independent speaker verification (TISV)~\cite{wan2017generalized}. In TDSV, the same text is used for enrollment and evaluation phases, while in TISV, there are no constraints on the enrollment or verification utterances, exposing a larger variability of phonemes and utterance durations~\cite{KINNUNEN201012, 10.1155/S1110865704310024}.
Combined with a keyword spotting (KWS) System, TDSV can be integrated into an intelligent personal assistant such as Apple Siri, Amazon Alexa, Google Now and Microsoft Cortana, where KWS and TDSV serve as a keyword voice-authenticated wake-up to enable the following voice interaction~\cite{7472652, zhang2017endtoend, 6854363}.

Before the deep neural networks era, the state-of-the-art SR method was the i-vector approach~\cite{inproceedingsdeha, 5545402, 4960564}.
Nowadays, DL methods are outperforming the former state-of-the-art methods in various fields of speaker recognition. 
However, in the context of TISV, the i-vector framework and its variants are still the state-of-the-art in some of the tasks~\cite{6853887, 10.1109/TASL.2010.2064307}. In NIST SRE12 and SRE16 and their post-evaluations, almost all leading systems are based on i-vectors~\cite{inproceedingsasd, nist-2012, nist-2016}. 
However, i-vector systems are prone to have performance degradation when short utterances are met in enrollment/evaluation phase~\cite{inproceedingsasd}.
Recently, DL-based, especially end-to-end, TISV has drawn more attention and many researchers have proposed different methods outperforming the i-vector/probabilistic linear discriminant analysis (PLDA) framework in various tasks.
According to the results reported in~\cite{inproceedingsasd, 7846260}, end-to-end DL systems achieved better performance compared to the baseline i-vector system~\cite{5545402}, especially for short utterances.
bidirectional long short-term memory units with a triplet loss achieved better performance in the “same/different” speaker detection experiment compared
with Bayesian Information Criterion and Gaussian Divergence~\cite{7953194}.



In this study, we discuss the generalized end-to-end (GE2E) DL-based technique proposed by~\cite{wan2017generalized} for TISV and examine various scenarios and parameters to further explain and evaluate the generality of the proposed generalized method.
The rest of our article is organized as follows.
In \cref{section:Methodology_new}, we present the end-to-end DL method and describe the utilized corpus and necessary data processing steps for TISV problem, as well as the training process.
\cref{section:Experiments} discusses different experiments performed to assess the performance of the end-to-end method. 
Finally, conclusions are stated in \cref{section:Discussion}.


\section{Methodology}
\label{section:Methodology_new}

An end-to-end system treats the entire system as a whole adaptable black box. The process of feature extraction and classifier training are achieved in parallel with an objective function that is consistent with the evaluation metric~\cite{Irum2019SpeakerVU}.
The main advantage of the generalized end-to-end training is that it enables us to process a large number of utterances at once, which greatly decreases the total training and convergence time. In this section, we first explain the proposed GE2E method \cite{wan2017generalized}.
Afterward, the necessary data pre-processing and preparation, training procedure, and configuration will be described.

\subsection{GE2E method}
\label{section:Methodology}

We select $N$ different speakers and fetch $M$ different utterances for every selected speaker to create a batch. Similar to~\cite{7472652}, the features $\mathbf{x}_{ji}$ extracted from each utterance will be fed to the network. The utilized network consists of 3 LSTM layers~\cite{10.1162/neco.1997.9.8.1735} followed by a linear projection layer in order to get to the final embedding vectors~\cite{articlesak}. The final embedding vector (d-vector) is the L2 normalization of the network output $f(\mathbf{x}_{ji};\mathbf{w})$ where $\mathbf{w}$ represents all parameters of the network,

\begin{equation}
    \mathbf{e}_{ji}=\frac{f(\mathbf{x}_{ji};\mathbf{w})}{\lVert f(\mathbf{x}_{ji};\mathbf{w}) \rVert_2},
\label{eq:L2norm}
\end{equation}
where $\mathbf{e}_{ji}$ represents the embedding vector of the $j$th speaker’s $i$th utterance. The centroid of the embedding vectors from the $j$th speaker
$[\mathbf{e}_{j1}, ... , \mathbf{e}_{jM}]$ $\mathbf{c}_{j}$ is defined as the arithmetic mean of the embedding vectors of the $j$th speaker. 


The similarity matrix $\mathbf{S}_{ji,k}$ is defined as the scaled cosine similarities between each embedding vector $\mathbf{e}_{ji}$ to all centroids $\mathbf{c}_{k}$ $(1 \leq j,k \leq N$, and $1 \leq i \leq M)$,

\begin{equation}
    \mathbf{S}_{ji,k}=w \cdot \cos(\mathbf{e}_{ji}, \mathbf{c}_{k})+b
\label{eq:cosine_mat},
\end{equation}
where $w$ and $b$ are learnable parameters. We constrain the weight to
be positive $w > 0$, because we want the similarity to be larger when
cosine similarity is larger.
Unlike most of the end-to-end methods, rather than a scalar value, GE2E builds a similarity matrix (\cref{eq:cosine_mat}) that defines the similarities between each $\mathbf{e}_{ji}$ and all centroids $\mathbf{c}_{k}$.


During training, we aim at maximizing the similarity of the embeddings representing the utterances of a particular speaker to centroid of embeddings of that speaker.
At the same time, we want to minimize the similarity of the embedding centroids of all other speakers. This general idea is borrowed from traditional methods, such as Linear Discriminant Analysis.

Furthermore, removing $\mathbf{e}_{ji}$ when computing the centroid of the true speaker makes training stable and helps avoid trivial solutions~\cite{wan2017generalized}. So, while we still take the arithmetic mean of the embedding vectors when calculating negative similarity (\ie $k \neq j$), we instead use the following when $k = j$,

\begin{equation}
    \mathbf{c}_{j}^{(-i)}=\frac{1}{M-1}\sum_{\substack{m=1 \\ m\neq i}}^{M}\mathbf{e}_{jm}.
\label{eq:centroid_without}
\end{equation}
Therefore, \cref{eq:cosine_mat} also becomes the following,

\begin{equation}
  \mathbf{S}_{ji,k} =
    \begin{cases}
      w \cdot \cos(\mathbf{e}_{ji}, \mathbf{c}_{j}^{(-i)})+b & \text{if $k=j$}\\
       w \cdot \cos(\mathbf{e}_{ji}, \mathbf{c}_{k})+b & \text{otherwise}.
    \end{cases}       
\end{equation}

Finally, we put a SoftMax on $\mathbf{S}_{ji,k}$ for $k = 1, ... , N$ that makes the output equal to 1 if $k = j$, otherwise makes the output equal to zero. Thus, the loss on each embedding vector $\mathbf{e}_{ji}$ could be defined as,

\begin{equation}
    L(\mathbf{e}_{ji})= -\mathbf{S}_{ji,j} + \log\sum_{k=1}^N\exp(\mathbf{S}_{ji,k}).
\label{eq:loss_single}
\end{equation}
This loss function means that we push each embedding vector close to its centroid and pull it away from all other centroids.


Finally, in order to calculate the final GE2E loss $L_G$, we have 2 options,
\begin{enumerate}
	\item According to ~\cite{wan2017generalized}, the GE2E loss $L_G$ is the sum of all losses over the similarity matrix ($1 \leq j \leq N$, and $1 \leq i \leq M$),
	
	\begin{equation}
   L_G(\mathbf{x};\mathbf{w})= L_G(\mathbf{S})=\sum_{j,i}L(\mathbf{e}_{ji}).
\label{eq:final_loss_sum}
\end{equation}

    \item the GE2E loss $L_G$ is the mean of all losses over the similarity matrix ($1 \leq j \leq N$, and $1 \leq i \leq M$),
    	\begin{equation}
   L_G(\mathbf{x};\mathbf{w})=\frac{1}{M\cdot N}\sum_{j,i}L(\mathbf{e}_{ji}).
\end{equation}

\end{enumerate}
Although both options eventually perform the same, We propose the option 2 as it is more consistent with changing the number of speakers per batch or utterances per speaker.


\subsection{Corpus and data pre-processing}
\label{section:data}

The dataset that we use for all the training, enrollment, and evaluation steps is the LibriSpeech dataset \cite{7178964}, which is derived from English audiobooks. The "train-clean-360" subset is used for training, while other subsets are used separately for enrollment and evaluation, in an open-set\footnote{In contrast to a closed-set scenario where evaluation and training speakers can overlap, in an open-set scenario, none of the speakers used in the training phase should be used for the evaluation.} manner. \cref{tab:data_stats} illustrates the statistics of the different subsets of the corpus. For each speaker in the "clean" training sets, the amount of speech is limited to 25 minutes, in order to avoid major imbalances in
per-speaker audio duration~\cite{7178964}. In the following, we describe the data pre-processing steps.

\begin{table}[t]
\centering
\caption{Statistics of the utilized subsets of the LibriSpeech corpus~\cite{7178964} in this study.}
\label{tab:data_stats}
\begin{tabular}{>{\centering\arraybackslash}m{1.8cm}>{\centering\arraybackslash}m{1.9cm}>{\centering\arraybackslash}m{1.1cm}>{\centering\arraybackslash}m{1.1cm}>{\centering\arraybackslash}m{1.1cm}}
\toprule
\textbf{subset name} & \textbf{subset duration [hours]} & \textbf{\# female speakers} & \textbf{\# male speakers} & \textbf{\# total speakers} \\
\midrule
dev-clean & \num{5.4} & \num{20} & \num{20} & \num{40} \\
test-clean & \num{5.4} & \num{20} & \num{20} & \num{40} \\
dev-other & \num{5.3} & \num{16} & \num{17} & \num{33} \\
test-other & \num{5.1} & \num{17} & \num{16} & \num{33} \\
train-clean-360 & \num{363.6} & \num{439} & \num{482} & \num{921} \\
\bottomrule
\end{tabular}
\end{table}

\subsubsection{Training data pre-processing}
After normalizing the volume of each utterance, we perform voice activity detection (VAD)~\cite{Ramirez07} with a maximum silence length of \SI{6}{ms} and a window length of \SI{30}{ms} followed by pruning the intervals with sound pressures below \SI{30}{db}. Therefore, we end up with smaller segments for each utterance, which are referred to as partial utterances~\cite{wan2017generalized}. We only select the partial utterances which are at least \SI{1.8}{s} long. 

Furthermore, the feature extraction process is the same as in~\cite{7178863}.
The partial utterances are first transformed into frames of width \SI{25}{ms}
with \SI{10}{ms} steps. Then we extract 40-dimensional log-Mel-filterbank
energies as the feature representation for each frame.

\subsubsection{Enrollment and evaluation data pre-processing}
Except for partial utterances, where we instead concatenate the resulting smaller segments of each utterance in order to have a single segment again for each utterance, the other steps remain the same here as in the training step.

\subsection{Training procedure}
\label{section:training}

We randomly choose $N$ speakers and randomly select $M$ pre-processed partial utterances for each speaker to construct a batch. 
In order to introduce more randomization, we randomly select a time length $t$ within $[140, 180]$ frames, and enforce that all partial utterances in that batch are of length $t$~\cite{wan2017generalized}. This means that partial utterances of different batches will have different lengths, but all the partial utterances in the same batch must be of the same length.

We use \num{768} hidden nodes and \num{256} dimensional embeddings for our network and optimize the model using the Adam~\cite{kingma2014adam} optimizer with a learning rate of $10^{-4}$. The network contains total of $12,134,656$ trainable parameters. Each batch consists of $N=16$ speakers and $M=5$ partial utterances per speaker, leading to \num{80} partial utterances per batch. The L2-norm of gradient is clipped at 3~\cite{Pascanu2012UnderstandingTE}, and the gradient scale for projection node in LSTM is set to 1. Furthermore, we initialize the scaling factor $(w, b)$ of the loss function with $(10, -5)$ and clamp $w$ to be larger than $10^{-6}$ in order to smooth the convergence. Moreover, the Xavier normal initialization~\cite{articlexavier} is applied to the network weights and the biases are initialized with zeros. 
\cref{alg:train_prep} and \cref{alg:train} explain the detailed training data pre-processing and training data batch preparation.

\begin{algorithm}[t]
 \For{all raw utterances}{
 $-$ normalize the volume\;
 $-$ perform VAD with $max\_silence\_length=6ms$ and $window\_length=30ms$\;
 $-$ prune the intervals with sound pressures below $30db$\;
 \For{all resulting intervals}{
   \uIf{interval's length $>$180 frames}{
    $-$ perform short-time Fourier transform (STFT) on the interval\;
    $-$ take the magnitude squared of the result\;
    $-$ transform to the Mel scale\;
    $-$ take the logarithm\;
   }}}
\caption{Proposed data pre-processing steps for training using the GE2E method for TISV.}
\label{alg:train_prep}
\end{algorithm}

\begin{algorithm}[t]
 \For{all training batches}{
  $-$ initialization: randomly choose an integer within $[140, 180]$ as the partial utterance length\;
 \For{all train speakers}{
 $-$ randomly choose $N$ speakers\;
 \For{all $N$ speakers}{
  $-$ randomly select $M$ partial utterances that are pre-processed according to \cref{alg:train_prep}\;
  \For{all $M$ partial utterances}{
    $-$ randomly segment an interval which has equal number of frames to the initialization step\;
   }}}}
\caption{Training iteration data preparation steps after data pre-processing.}
\label{alg:train}
\end{algorithm}



\section{Experiments}
\label{section:Experiments}

In order to assess the performance of the proposed method in \cref{section:Methodology}, we compare the evaluation results with a baseline method (\cf\cref{section:baseline}) and also discuss various experiments in this section.
Before getting to the experiments, we first need to clarify the process of obtaining the d-vectors for enrollment and evaluation utterances and also explain the utilized evaluation and quantitative analysis approach.


\subsection{Enrollment and evaluation d-vectors}
\label{section:dvector}

For the sake of convenience and time, we first feed all the available pre-processed enrollment and evaluation utterances to the trained network (\cf\cref{section:training}) and store the resulting d-vectors. Subsequently, we could easily load them to perform enrollment and evaluation processes for various experiments.


For every utterance we apply a sliding window of fixed size $(140 + 180)/2 = 160$ frames with \SI{50}{\percent} overlap.
We compute the d-vector for each window. The final utterance-wise
d-vector is generated by first L2 normalizing the window-wise d-vectors, and then followed by taking the element-wise average~\cite{wan2017generalized}.
The detailed descriptions of the enrollment and evaluation data pre-processing and preparing for d-vector creation are given by \cref{alg:test_prep} and \cref{alg:test}.

\begin{algorithm}[t]
 \For{all raw utterances}{
 $-$ normalize the volume\;
 $-$ perform VAD with $max\_silence\_length=6ms$ and $window\_length=30ms$\;
 $-$ prune the intervals with sound pressures below $30db$\;
 \For{all resulting intervals}{
   \uIf{interval's length $<$ 180 frames}{
   $-$ drop the interval\;
   }}
  $-$ concatenate the remaining intervals\;
  $-$ perform STFT on the concatenated utterance\;
  $-$ take the magnitude squared of the result\;
  $-$ transform to the Mel scale\;
  $-$ take the logarithm\;
   }
\caption{Proposed data pre-processing steps for enrollment and evaluation using the GE2E method for TISV. STFT stands for short-time Fourier transform.}
\label{alg:test_prep}
\end{algorithm}

\begin{algorithm}[t]
 \For{all enrollment and evaluation speakers}{
  \For{all pre-processed utterances}{
  
   $-$ initialization: set the starting time frame of the window $t=0$\;
   \While {not reached the end of the utterance}{

    $-$ select the interval within $[t, t+160]$ frames of the utterance\;
    $-$ feed the selected utterance to the trained network to obtain the corresponding d-vector\;
    $-$ L2-normalize the d-vector\;
    $-$ $t=t+80$\;
   }
   $-$ Perform element-wise average of the L2-normalized d-vectors to obtain the final utterance d-vector\;
   }}
\caption{Enrollment and evaluation data preparation and d-vector creation after data pre-processing.}
\label{alg:test}
\end{algorithm}


\subsection{Quantitative analysis approach}
\label{section:metrics}

After creating d-vectors, we can start with evaluating the system. we use a threshold-based binary classification method in this stage, where we first need to create a speaker reference model for each speaker to be evaluated, \ie the enrollment step.
In the next step, we calculate the similarity between the unknown test utterance d-vector and the already built speaker model d-vector.
The similarity metric, which we use here, is the cosine distance score, which is the normalized dot product of the speaker model and the test d-vector,

\begin{equation}
    \cos(\mathbf{e}_{ji} , \mathbf{c}_{k})=\frac{\mathbf{e}_{ji} \cdot \mathbf{c}_{k}}{\lVert \mathbf{e}_{ji} \rVert \cdot \lVert \mathbf{c}_{k} \rVert}.
\label{eq:cosine}
\end{equation}
The higher the similarity score between $\mathbf{e}_{ji}$ and $\mathbf{c}_{k}$ is, the more similar they are.

The metric which we use for the evaluation of the performance of our speaker verification system is referred to as equal error rate (EER), which is used to predetermine the threshold values for its false acceptance rate (FAR) and its false rejection rate (FRR)~\cite{vanLeeuwen2007, Hansen2015SpeakerRB}.
It searches for a threshold for similarity scores where the proportion of genuine utterances which are classified as imposter (FRR) is equal to the proportion of imposters classified as genuine (FAR).

The overall FAR, FRR, and EER are calculated according to \cref{eq:FAR}, \cref{eq:FRR}, and \cref{eq:EER}, respectively. True acceptance (TA), true rejection (TR), false acceptance (FA), and false rejection (FR) values are used for the calculations. Note that, since the FAR and FRR curves are monotonic, there is only one point where the FAR has the same value as the FRR. 

\begin{equation}
    FAR = \frac{FA}{FA+TR}
\label{eq:FAR}
\end{equation}

\begin{equation}
    FRR = \frac{FR}{FR+TA}
\label{eq:FRR}
\end{equation}

\begin{equation}
    EER = \frac{FAR+FRR}{2},  \:\:\:\:\:\:\:\text{if} \:\:\: FAR = FRR
\label{eq:EER}
\end{equation}

\subsection{The baseline system}
\label{section:baseline}

The baseline is a standard i-vector system proposed by~\cite{5545402}. \cref{tab:ivector} shows the evaluation results on "dev-clean" and "test-clean" subsets. The experiments are performed for three cases with different i-vector dimensions and different Gaussian mixture model (GMM) components, with random data split and simple thresholding. Each positive sample is tested against 20 negative samples, and 20 different positive samples are tested per speaker. From \cref{tab:ivector}, we can already observe that the EER results are quite high with the baseline system.

\begin{table}[t]
\centering
\caption{The evaluation results using the baseline i-vector method~\cite{5545402} with random data split and simple thresholding on "dev-clean" and "test-clean" datasets. Each positive sample is tested against 20 negative samples. Furthermore, 20 different positive samples are tested per speaker. Columns one and two show the i-vector dimensionality and number of GMM elements, respectively.}
\label{tab:ivector}
\begin{tabular}{cccc}
\toprule
\textbf{i-Vec dimension} & \textbf{GMM elements} & \textbf{dev-clean EER} & \textbf{test-clean EER} \\
\midrule
\num{600} & \num{1024} & \SI{16.65}{\percent} & \SI{18.58}{\percent} \\
\num{400} & \num{512} &\SI{17.80}{\percent} & \SI{17.70}{\percent} \\
\num{300} & \num{256} & \SI{18.86}{\percent} & \SI{16.90}{\percent} \\
\bottomrule
\end{tabular}
\end{table}

\subsection{Performance by number of enrollment utterances}
\label{section:utternum}


Usually in SV, there are multiple enrollment utterances for each speaker in order to build a robust speaker model.
The observed EER is only an approximation of the system's true EER. Consequently, we repeat the enrollment and evaluation processes for $1K$ iterations and average the results to make up for the aforementioned problem.
Moreover, while $M$ utterances for every speaker should be randomly selected in order to construct a batch for processing, we choose $N$ to be equal to the number of all the available speakers in the test set in order to  further reduce the randomization imposed by sampling.

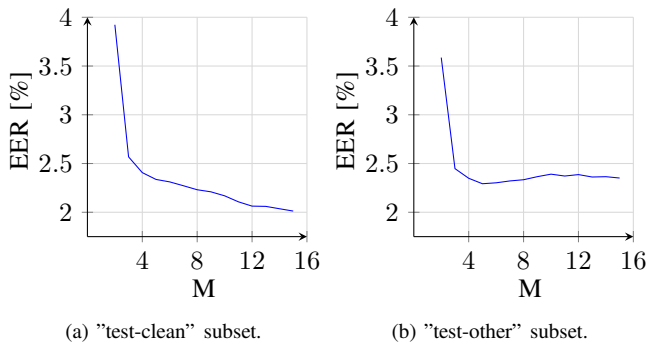
\begin{figure}[t]
\begin{subfigure}{.49\linewidth}
\begin{tikzpicture}
\begin{axis}[
                 grid=both,
                 grid style={solid,gray!30!white},
                 axis lines=middle,
                 xlabel={M},
                 ylabel={EER [\%]},
                 width=4.5cm,
                 height=4.5cm,
                 xtick={0,4,...,16},
                 ytick={0,0.5,...,4.5},
                 xmin=0,
                 xmax=16,
                 ymin=1.75,
                 ymax=4.0,
                 xlabel={M},
                 x label style={at={(axis description cs:0.53,-0.15)},anchor=north},
                 y label style={at={(axis description cs:-0.21,0.53)},rotate=90,anchor=south},]
\addplot[blue] table [x=Step, y=Value, col sep=comma] {figs/enrol_utter_num_test_clean.csv};
\end{axis}
\end{tikzpicture}
\caption{"test-clean" subset.}
\end{subfigure}%
\begin{subfigure}{.49\linewidth}
\begin{tikzpicture}
\begin{axis}[
                 grid=both,
                 grid style={solid,gray!30!white},
                 axis lines=middle,
                 xlabel={M},
                 ylabel={EER [\%]},
                 width=4.5cm,
                 height=4.5cm,
                 xtick={0,4,...,16},
                 ytick={0,0.5,...,4.5},
                 xmin=0,
                 xmax=16,
                 ymin=1.75,
                 ymax=4.0,
                 xlabel={M},
                 x label style={at={(axis description cs:0.53,-0.15)},anchor=north},
                 y label style={at={(axis description cs:-0.21,0.53)},rotate=90,anchor=south},]
\addplot[blue] table [x=Step, y=Value, col sep=comma] {figs/enrol_utter_num_test_other.csv};
\end{axis}
\end{tikzpicture}
\caption{"test-other" subset.}
\end{subfigure}
\caption{Average EER [\%] over $1K$ test iterations for different numbers $M$ of enrollment d-vectors per speaker, separately on the (a) "test-clean", and (b) "test-other", subsets of LibriSpeech. The minimum possible $M$ is 2, as we are removing the utterance itself when calculating centroids based on \cref{eq:centroid_without}.}
\label{fig:enrol_utter_num}
\end{figure}

\cref{fig:enrol_utter_num} and \cref{tab:enrol_utter_num} show the average EER over $1K$ test iterations for different numbers $M$ of enrollment d-vectors per speaker, separately on different subsets of LibriSpeech \cite{7178964}. Note that the minimum possible $M$ is 2, as we are averaging over the enrollment d-vectors in order to get the speaker models, while removing the utterance itself when calculating centroids based on \cref{eq:centroid_without}. Also, in every test iteration, we select $2M$ utterances per speaker and split them in half for the enrollment and evaluation steps. As we can see, the choice of $M$ is the most decisive for the lower values. Moreover, the curve is monotonically decreasing for the clean environment, while for the noisy "test-other" set, increasing $M$ does not make improvements for higher values.

\begin{table}[t]
\centering
\caption{Average EER [\%] over $1K$ test iterations for different numbers of enrollment d-vectors per speaker for different evaluation subsets. The last column shows the average over all $M$ for $M=2,3,...,15$.}
\label{tab:enrol_utter_num}
\begin{tabular}{>{\centering\arraybackslash}m{1.5cm}>{\centering\arraybackslash}m{0.45cm}>{\centering\arraybackslash}m{0.45cm}>{\centering\arraybackslash}m{0.45cm}>{\centering\arraybackslash}m{0.45cm}>{\centering\arraybackslash}m{0.45cm}>{\centering\arraybackslash}m{0.45cm}>{\centering\arraybackslash}m{0.45cm}}
\toprule
\textbf{\# enroll utts} & \textbf{2} & \textbf{3} & \textbf{4} & \textbf{7} & \textbf{10} & \textbf{15} & Avg \\
\midrule
test-clean & \num{3.92} & \num{2.57} & \num{2.41} & \num{2.27} & \num{2.17} & \num{2.01} & \num{2.34}\\
test-other & \num{3.59} & \num{2.45} & \num{2.35} & \num{2.32} & \num{2.39} & \num{2.35} & \num{2.44}\\
dev-clean & \num{3.21} & \num{1.94} & \num{1.94} & \num{1.89} & \num{1.89} & \num{1.81} & \num{1.97}\\
\bottomrule
\end{tabular}
\end{table}

\subsection{Performance on the test set}
\label{section:test}

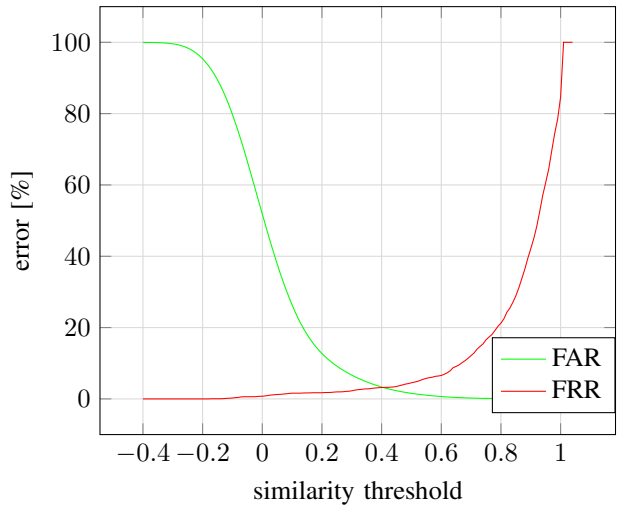
\begin{figure}[t]
\begin{tikzpicture}
\centering
\begin{axis}[
                 grid=both,
                 grid style={solid,gray!30!white},
                 xlabel={similarity threshold},
                 ylabel={error [\%]},
                ]
\addplot[green] table [x=Step, y=Value, col sep=comma] {figs/FAR.csv};
\addplot[red] table [x=Step, y=Value, col sep=comma] {figs/FRR.csv};
\legend{FAR, FRR}
\end{axis}
\end{tikzpicture}
\caption{The FAR and FRR values over different similarity thresholds for $M=2$ averaged over $1K$ test iterations on "dev-clean" set. The EER is equal to the value which lies at the intersection point of two curves, which approximately equals \num{0.42} here.}
\label{fig:test}
\end{figure}

In this experiment, we first perform the enrollment and evaluation tasks on the "dev-clean" set for $M=2$ and fix the obtained average threshold and use to perform enrollment and verification on the "test-clean" and "test-other" sets. \cref{fig:test} illustrates the FAR and FRR values over different similarity thresholds. The EER is the intersection between two curves. \cref{tab:fixed_thres} also shows the evaluation results on the test sets tested with the fixed threshold obtained from "dev-clean". Furthermore, \cref{tab:diff_epoch} shows the evaluation results on the "test-clean" using the model trained after different epochs, which proves how fast the network converges.

\begin{table}[t]
\centering
\caption{The evaluation results for $M=2$ averaged over $1K$ test iterations on the test sets with a fixed threshold of $0.42$ obtained from "dev-clean".}
\label{tab:fixed_thres}
\begin{tabular}{>{\centering\arraybackslash}m{2.9cm}>{\centering\arraybackslash}m{0.6cm}>{\centering\arraybackslash}m{0.6cm}>{\centering\arraybackslash}m{0.6cm}}
\toprule
\textbf{Evaluation metrics[\%]} & EER & FAR & FRR \\
\midrule
test-clean & \num{3.85} & \num{3.68} & \num{4.03} \\
test-other & \num{3.66} & \num{2.89} & \num{4.43} \\
\bottomrule
\end{tabular}
\end{table}

\begin{table}[t]
\centering
\caption{Evaluation results for $M=2$ averaged over $1K$ test iterations on the "test-clean" suubset using the model trained at different stages. Every column shows the number of epochs used to train the model. The last column shows the final model.}
\label{tab:diff_epoch}
\begin{tabular}{cccccc}
\toprule
\textbf{\# train epochs} & \textbf{40} & \textbf{80} & \textbf{120} & \textbf{160} & final \\
\midrule
EER[\%] & \num{6.12} & \num{5.08} & \num{4.42} & \num{4.44} & \num{3.96} \\
FAR[\%] & \num{6.12} & \num{5.10} & \num{4.44} & \num{4.44} & \num{3.97} \\
FRR[\%] & \num{6.12} & \num{5.06} & \num{4.40} & \num{4.44} & \num{3.95} \\
\bottomrule
\end{tabular}
\end{table}

\subsection{Performance by test utterance duration}
\label{section:utterdur}


Even though the state-of-the-art DL methods have outperformed most of the traditional methods in various speaker recognition task and shown outstanding results, text-independent speaker verification is still a challenging problem when it comes to short-length utterances. In this experiment, we evaluate the performance of our method separately for short and long utterances. We consider an utterance short when its duration is less than 4 seconds and long when its duration is longer than $4s$. \cref{tab:data_stats_duration} shows the number of utilized short and long utterances available per subset. As shown in \cref{tab:short_dur}, performance drops significantly by \SI{59}{\percent} when only considering the short-length utterances compared to the unconstrained case for the "test-clean" subset.

\begin{table}[t]
\centering
\caption{Number of utterances in LibriSpeech test sets based on utterance duration. The last column also shows the total number of speakers per subset.}
\label{tab:data_stats_duration}
\begin{tabular}{ccccc}
\toprule
\textbf{subset} & \textbf{$\leq 4s$} & \textbf{$> 4s$} & total & \# speakers \\
\midrule
dev-clean & \num{765} & \num{1938} & \num{2703} & \num{40} \\
test-clean & \num{772} & \num{1848} & \num{2620} & \num{40} \\
test-other & \num{1022} & \num{1917} & \num{2939} & \num{33} \\
\bottomrule
\end{tabular}
\end{table}

\begin{table}[t]
\centering
\caption{EER [\%] results for $M=2$ for different utterance lengths averaged over 1K test iterations on the test sets. First column shows the results for short utterances, while the second column states the results for long utterances. Finally, the last column shows the results without taking the utterance duration into consideration.}
\label{tab:short_dur}
\begin{tabular}{cccc}
\toprule
\textbf{subset} & \textbf{$\leq 4s$} & \textbf{$> 4s$} & total \\
\midrule
dev-clean & \num{3.40} & \num{3.01} & \num{3.32} \\
test-clean & \num{6.32} & \num{3.51} & \num{3.98} \\
test-other & \num{5.90} & \num{3.58} & \num{3.61} \\
\bottomrule
\end{tabular}
\end{table}



\section{Conclusions}
\label{section:Discussion}

In this study, we investigated the GE2E method proposed in~\cite{wan2017generalized} for text-independent speaker verification.
The results verified the advantages of this method compared to the conventional baseline systems. We observed that GE2E training was about $3\times$ faster than the conventional speaker verification systems and converges very fast, while it is one of the few DL-based TISV methods that outperforms the baseline system.
Furthermore, even though short-length utterances are more difficult to predict, we showed that the proposed method is flexible in utterance duration and still works well for short-duration data.
Moreover, as it was expected, we observed that the choice of number of  utterances per enrollment speaker ($M$) is an effective factor on the final EER value, where with increasing $M$, the resulting EER would be lower. Also, we observed that the proposed method generalizes very fast in this issue and shows great performance with already a few number of enrollment utterances per speaker. 



\bibliographystyle{IEEEtran}
\bibliography{Internship}

\end{document}